
%
%
\magnification=1200
\vsize=7.5in
\hsize=5in
\pageno=1
\tolerance=10000
\null
\vskip 1.0in
\centerline{\bf ABSTRACT}
\vskip 1.0in
\baselineskip 24pt plus 4pt minus 4pt
We investigate the magnetic field behaviour of an antiferromagnetic Heisenberg
spin-1 chain with the most general single-ion anisotropy.
We discuss the regime in which the magnetic field is below the transition
value.
The splitting of
the Haldane triplet is obtained as a function of a field applied in an
arbitrary orientation by means of a Lancz\H os exact diagonalization
of  chains of up to 16 spins. Our results are nicely summarized in terms
of a first-order perturbation theory. We explain various level crossings
that occur by the existence of discrete symmetries. A discussion is given
of the electron spin resonance and neutron scattering experiments on the
compound Ni(C$_2$H$_8$N$_2$)$_2$NO$_2$ClO$_4$ (NENP).
\vfill
\eject
\magnification=1200
\baselineskip 24pt plus 4pt minus 4pt
\noindent{\bf I. INTRODUCTION}
\medskip
There is good evidence that the integer-spin Heisenberg antiferromagnetic
chains have a gap, as suggested by Haldane$^{1}$. This is established on
the theoretical side by various techniques including finite-size calculations
as well as field theoretic arguments$^2$. On the experimental side the first
evidence$^3$ came from CsNiCl$_3$: neutron scattering (NS) revealed an
excitation
gap. Since this compound is only moderately one-dimensional, parasitic
three-dimensional effects complicate the picture. On the other hand,
the compound$^4$
Ni(C$_2$H$_8$N$_2$)$_2$NO$_2$ClO$_4$, abbreviated NENP, is much more
one-dimensional and remains in a magnetically disordered state even at very low
temperature. Zero-field neutron scattering experiments$^5$ have clearly
shown the existence of the Haldane triplet split by easy-plane anisotropy.
In addition, a number of other measurements have been performed$^{5-8}$ in a
magnetic field: susceptibility, high-field magnetization, and neutron
scattering in a finite field. The application of a magnetic field leads to a
Zeeman splitting of the Haldane triplet and one member of this triplet crosses
the
ground state at a critical value H$_c$ that depends on the field orientation.
This
is clearly seen in NS experiments where all members of the triplet
can be followed individually$^9$. Experiments using electron spin resonance
technique (ESR) are in excellent agreement$^{10}$ with NS
as is the case for far-infrared spectroscopy measurements$^{11}$ and there is
at
the present time a satisfactory picture of the behaviour of NENP from the
experimental point of view.

On the theoretical side, effective  quantum field theories have been used
to predict the magnetic field behaviour of the spin-1 chain Heisenberg
Hamiltonian  that models
the magnetic properties of NENP. In fact the original work of Haldane showed
that,
in the large integer spin limit, the antiferromagnetic spin chain is described
in
the low-energy limit by an O(3) nonlinear sigma model. This nontrivial field
theory
is difficult to study and, based on the large-N limit of the O(N) model, it
has been
suggested that a simple theory with three massive bosonic fields might be an
appropriate effective theory of the spin-1 chain. It is then possible to obtain
the behaviour$^{12,13}$ of the system in a magnetic field.

Another possibility has been suggested starting
from an integrable chain$^{14}$. It is known that the spin-1 Hamiltonian
${\cal H}_0 =\sum_i {\bf S}_i \cdot {\bf S}_{i+1} - ({\bf S}_i \cdot {\bf
S}_{i+1})^2$ is solvable by the Bethe Ansatz technique$^{15}$ and leads to a
massless theory. This massless theory is characterized by an SU(2)$_{k=2}$
symmetry
and can be realized by three (massless) Majorana fermions. As one perturbs the
Hamiltonian $H_0$ towards the pure Heisenberg Hamiltonian without biquadratic
coupling, it is natural in the framework of the Haldane conjecture to expect
that
these fermions become massive. One can then use a theory of three free massive
Majorana fermions to approximate the Heisenberg chain. These two
theoretical approaches
are not in complete agreement and it is thus interesting to have results
of a completely different nature, based on numerical studies of finite
chain diagonalization.

In this paper, we present the results of our study of the field behaviour
of a spin-1 chain including realistic single-ion anisotropies of the most
general
kind. We diagonalize by means of a Lancz\H os algorithm chains of up to
16 spins under a magnetic field applied in various positions. Our findings are
neatly summarized by a simple perturbation calculation that may be used as a
practical tool to obtain the field behaviour of an Haldane magnet.
When the applied field become strong enough there is a phase transition
towards a magnetically ordered phase$^{12,13}$. In this paper
we will restrict ourselves to the singlet phase where the Haldane gap
is not destroyed.
We give a discussion of the various level crossings that may or may not appear
depending on the field orientation with respect to the symmetry axis of the
crystal. Section II contains the treatment of the magnetic field
as a perturbation.
Section III explains the Lancz\H os results as well as their relationships with
the perturbative expansion. Section IV contains our conclusions and a
discussion of NENP experiments.
\bigskip
\noindent{\bf II. THE PERTURBATIVE RESULTS}
\medskip
We focus on the microscopic Hamiltonian of a spin-1 antiferromagnetic chain
in an applied field and single-ion anisotropy:
$$
{\cal H}= J\sum_i {\bf S}_i \cdot {\bf S}_{i+1} + D(S^z_i )^2 + E[ (S^x_i )^2 -
(S^y_i )^2 ] - {\bf H}\cdot {\bf S}_i .
\eqno(II-1)
$$
Here ${\bf S}_i$ are quantum spins S=1 and
we include the Bohr magneton and the Land\'e $g$ factors in the definition
of the magnetic field. The exchange coupling $J$ is taken to be positive
i.e. antiferromagnetic. Below we set $J=1$.
The D and E terms in Eq.(II-1) parametrize
the most general single-ion anisotropy. We work with periodic boundary
conditions.
In zero field and in the absence of
anisotropy ($D=E=0$) the Hamiltonian (II-1) is invariant under the full SU(2)
rotation group. If the $D$ term is nonzero then the symmetry is broken  to a
residual U(1) subgroup of rotations around the $Z$ axis. If, in addition to
$D$,
there is some further in-plane anisotropy  (nonzero E) even this U(1) is
broken.
However in this case there are still discrete remnants of the initial SU(2):
the
system is invariant under $\pi$ rotations around the coordinate axis $X, Y, Z$.
These symmetry operations are denoted by $R^{\pi}_x ,R^{\pi}_y ,R^{\pi}_z$
in this article.

If we now add a magnetic field, there is further symmetry reduction:
when $\bf H$ is in a generic position with respect to the coordinate axis,
even the discrete operations $R^{\pi}_x ,R^{\pi}_y ,R^{\pi}_z$ are lost.
However, there are some special orientations of {\bf H} that retain
discrete symmetries: if {\bf H} lies along the $\alpha$ axis the symmetry
$R^{\pi}_{\alpha}$ is conserved. This corresponds simply to rotation
around the magnetic field axis. But there are also discrete symmetries when
the magnetic field lies in one of the planes $(X, Y)$, $(Y, Z)$ or $(X, Z)$.
This is seen easily when the Y-component of {\bf H} is zero: the Hamiltonian
expressed in the basis of eigenstates of $S^z_i$ is then a {\it real} symmetric
matrix.  Complex conjugation is thus a discrete symmetry. In fact
Complex conjugation simply changes the sign of the operators $S^y_i$
and is a symmetry when $H_y =0$ (in this particular basis).
When the magnetic field is in another
plane the system is still invariant under a combined operation involving
complex conjugation and a rotation $R^{\pi}_{\alpha}$. If, for example, $H_z
=0$
then complex conjugation will transform $H_y S^y_i $ into $-H_y S^y_i $
and then one uses $R^{\pi}_{x}$ to come back to the original Hamiltonian.
These extra symmetries will explain the various
level crossings that are found in our exact diagonalization studies reported in
section III.

We now discuss the application of perturbation theory with respect to the
D, E, and {\bf H} terms in Hamiltonian (II-1). We write Eq.(II-1) as:
$$
{\cal H}= {\cal H}_0 + {\cal H}_I ,
$$
$$
{\cal H}_0 = \sum_i {\bf S}_i \cdot {\bf S}_{i+1},
\eqno(II-2)
$$
$$
{\cal H}_I =\sum_i D(S^z_i )^2 + E[ (S^x_i )^2 -
(S^y_i )^2 ] - {\bf H}\cdot {\bf S}_i .
$$
The Hamiltonian ${\cal H}_0$ has full rotational symmetry and its levels can
thus be classified according to their spin. We focus on the effect of the
perturbation on a singlet and a triplet state. We know that the ground state of
${\cal H}_0$ is in fact a singlet with chain momentum $K=0$
and the first excited state is a
triplet with momentum $K=\pi $. Our statements about the effect of
the perturbation
are general since they are dictated by the Wigner-Eckhardt theorem.

Let us first discuss the case of a singlet state $|0\rangle$. Due to
complete isotropy, the following equalities hold:
$$
\langle 0 | \sum_i ( S^x_i )^2 | 0 \rangle =
\langle 0 | \sum_i ( S^y_i )^2 | 0 \rangle =
\langle 0 | \sum_i ( S^z_i )^2 | 0 \rangle = {1\over 3} NS(S+1)={2\over 3} N.
\eqno(II-3)
$$
In addition the vector $\langle 0 | \sum_i  {\bf S}_i  | 0 \rangle$ is zero.
We thus obtain the first order shift of the singlet energy:
$$
E^{(1)} = E_0 + D\langle 0 | \sum_i ( S^z_i )^2 | 0 \rangle
 = E_0 + {2\over 3} D N.
\eqno(II-4)
$$
We note that there is no effect coming from the in-plane E-term or ${\bf H}$.
We now discuss the triplet splitting in zero field. In the standard basis
the triplet states are noted $|1m\rangle$, $m=-1,0,+1$. Perturbation theory
involves matrix elements of the following operator:
$$
{\cal O}^{\alpha\beta}=\sum_i S_i^{\alpha} S_i^{\beta}
-{2\over 3}\delta^{\alpha\beta}N.
\eqno(II-5)
$$
It is a spin-2 irreducible tensor operator since it transforms as a traceless
symmetric tensor. Its standard components are noted ${\cal O}^{(2M)}$,
$M=-2,\dots ,+2$.
The Wigner-Eckhardt theorem implies that the matrix
elements in a triplet state of such an operator are related by:
$$
\langle 1 m|{\cal O}^{(2M)}|1 m^{\prime}\rangle =
C \cdot \langle 1 m|21Mm^{\prime}\rangle .
\eqno(II-6)
$$
In this equation $\langle 1 m|21Mm^{\prime}\rangle$ is the Clebsch-Gordan
coefficient coupling two spin-1 states to a spin-2 state. Thus perturbation
theory
is characterized by a single coefficient $C$ (to first order). This
is easily found when working out matrix elements in the canonical basis
$|x\rangle ,|y\rangle ,|z\rangle $ rather than using the standard basis.
We consider the matrix elements:
$$
\langle \alpha | \sum_i (S_i^{\beta})^2 | \gamma \rangle .
\eqno(II-7)
$$
Application of a $\pi$ rotation shows that this element is zero if
$\alpha$ and $\gamma$ are distinct. In this case there are only two
different matrix elements. We can choose:
$$
\langle x | \sum_i (S_i^{z})^2 | x \rangle (\equiv a)
\quad {\rm and} \quad
\langle x | \sum_i (S_i^{x})^2 | x \rangle (\equiv b) .
\eqno(II-8)
$$
In addition one notes that $b+2a=NS(S+1)=2N$. The perturbation
${\cal H}_S =\sum_i
D(S^z_i )^2 + E[ (S^x_i )^2 - (S^y_i )^2 ]$ in the canonical basis for the
triplet can thus be written as:
$$
{\cal H}_S = \left(
\matrix{aD+(2N-3a)E&0&0\cr 0&aD+(3a-2N)E&0\cr 0&0&2(N-a)D\cr}\right) .
\eqno(II-9)
$$
As dictated by Wigner-Eckhardt theorem there is only one coefficient that
characterizes the perturbation. Subtracting the ground state energy one
is led to the following gap values:
$$
\eqalign{&\Delta_x = \Delta -\kappa D +3 \kappa E\cr
&\Delta_y = \Delta -\kappa D -3 \kappa E\cr
&\Delta_z = \Delta +2\kappa D .\cr}
\eqno(II-10)
$$
In this equation $\Delta$ is the unperturbed triplet-singlet gap and
we note $\kappa =2N/3-a$. The coefficient $a$ is extensive but the difference
$2N/3-a$ is finite in the thermodynamic limit since $\kappa$ appears
in gap values.
The value of $\kappa$ is not dictated by rotational
symmetry of course and its numerical value depends for example of the moment of
the triplet. When $E=0$ this splitting has been studied in detail$^{16,17}$ for
the Haldane triplet with chain momentum $\pi$ which is the lowest lying
triplet.
We note that already for $D\approx 0.2 J$ there are deviations from the
previous perturbation theory: the slopes with D of the two gaps are found to
be:
$$
\eqalign{&\Delta_x =\Delta_y = \Delta -0.57 D \cr
&\Delta_z = \Delta +1.41 D .\cr}
\eqno(II-11)
$$
This is the best linear fit of the Lancz\H os results between $D=0.10 J$
and $D=0.25 J$. The slope ratio is already slightly different from 2 as given
by
Eq.(II-10).  The curvature of the gaps as functions of D is clearly seen in the
data  of ref.[17]. In fact a fit including quadratic terms in D of the same
data leads to the following result:
$$
\eqalign{&\Delta_x =\Delta_y = \Delta -0.668 D + 0.269 D^2\cr
&\Delta_z = \Delta +1.357 D +  0.135 D^2 .\cr}
\eqno(II-12)
$$
The first-order terms satisfy the perturbative results (II-10) and we clearly
see the deviation from first-order. For practical purposs it is simpler to use
the fits (II-11).

We now add the magnetic field to the perturbative treatment. The ground state,
being a singlet, is not affected at first order and only the triplet changes.
The matrix of the perturbation in the canonical basis is then:
$$
{\cal H}_S - {\bf H}\cdot {\bf S} = \left(
\matrix{p_x & iH_z & -iH_y \cr -iH_z & p_y & iH_x \cr
iH_y & -iH_x & p_z \cr}\right) .
\eqno(II-13)
$$
The quantities $p_{\alpha}$ denote the diagonal matrix elements of Eq.(II-9).
Shifting the origin of the energies it is convenient to set $p_{\alpha}
= \Delta_{\alpha}$ and then the eigenenergies of (II-13) are directly the gaps.
In section III we show that the resulting values are always extremely
close of the Lancz\H os results when the magnetic field is below the critical
value. Diagonalization of (II-13) does not lead to compact formulas except
when the field lies along one of the symmetry axis. If ${\bf H}$ lies
along the direction $\gamma$ we denote its only nonzero component by
$H_{\gamma}$.
The eigenvalue $\Delta_{\gamma}$ is unperturbed and the two other eigenvalues
are given by:
$$
\Delta^{\pm}={1\over 2}\left[ \Delta_{\alpha} + \Delta_{\beta} \pm
\left[\left(\Delta_{\alpha}-\Delta_{\beta}\right)^2 +4 H_{\gamma}^2
\right]^{1/2}\right] .
\eqno(II-14)
$$
Here $\alpha$ and $\beta$ are the two other coordinates. There is hyperbolic
repulsion of the two gaps $\Delta^+$ and $\Delta^-$. The smallest gap
$\Delta^-$
goes to zero for a critical value $H_c^2 = \Delta_{\alpha}\Delta_{\beta}$.
At large fields the asymptotic behaviour of the gaps is linear.
Of course the crossing of one member of the triplet with the ground state
signals a phase transition$^{12,13}$ beyond which we do not expect to
gain something from a simple perturbative calculation since other states
with higher spins also cross the ground state above $H_c$.

It is interesting to note that this hyperbolic behaviour (II-14) is exactly
what is found in the fermionic effective theory$^{14}$ for states with
$K=\pi$. When the field becomes large and parallel to a coordinate axis
the eigenstates of (II-13) take a simple form:
 they are given by $|\alpha\rangle \pm i|\beta\rangle$. We note that
the vanishing of $\Delta^-$ at $H_c$ occurs linearly contrary to the
free boson prediction.

If the gap value $\Delta_{\gamma}$ (which does not move with the field)
lies above or below the two gaps $\Delta_{\alpha}$ and $\Delta_{\beta}$
there is in general a crossing of levels between  one of the gaps
$\Delta^{\pm}$ and $\Delta_{\gamma}$ before the critical field.
As shown in sect. III since they behave differently under exact
discrete symmetries of the system, we expect that these crossings
will survive beyond perturbation theory. If however the magnetic field is
no longer in a high symmetry position these symmetries are broken and
one should see only avoided crossings.
\bigskip
\noindent{\bf III. LANCZOS RESULTS}
\medskip
We have performed a Lancz\H os study of the Hamiltonian (II-1) on chains
of lengths N=4,6,8,10,12,14,16.
For a generic orientation of the magnetic field, there are no symmetries
available apart translational symmetry to reduce the size of the problem.
Thus many iterations were required to get the first few excited levels.
The energies of the ground state (in the subspace $K=0$) and the three
low-lying levels (in the subspace $K=\pi$) have been obtained with a typical
precision of 10$^{-6}$. For a generic orientation of the magnetic field,
the size of the complex Hermitian matrix $\cal H$, 3$^N$, is reduced by
translational symmetry to $\approx 3^N / N$. The size of the Hilbert space
for N=16 is $\approx 1.3\times 10^6$. On one Cray-2 processor, acting
with $\cal H$ on a vector takes about 17 seconds and the precision
of 10$^{-6}$ is reached for $\approx 60$ iterations.

We have followed the triplet splitting as a function of the
E-term. Neutron scattering experiments$^{18}$ have shown that there is
a small splitting of the two low-lying modes in the case of NENP:
$\Delta_x\approx 1.05$ meV and $\Delta_y\approx 1.25$ meV. This means that
the E-term is much smaller than the D-term in Eq.(II-1) since $\Delta_z\approx
2.5
meV$. We treat it
perturbatively but keep the D-term in the unperturbed Hamiltonian to be solved
by
Lancz\H os technique. When E=0 the Haldane triplet is split in a high-energy
singlet and a doublet. First-order perturbation theory for the E-term requires
its
matrix elements in the subspace spanned by the doublet. We find a linear
splitting:
$$
\eqalign{
&\Delta_x = \Delta_0 (D) + \kappa_0 (D) E\cr
&\Delta_y = \Delta_0 (D) - \kappa_0 (D) E\cr}
\eqno(III-1)
$$
Here $\Delta_0 (D)$ is the doublet gap when E=0. It is known by eq.(II-11).
The constant $\kappa_0 (D)$ has been computed for all lattice sizes.
Its values as a function of $D$ are given in fig.1.
We have used the Shanks algorithm$^{16}$ to obtain an estimate of the
thermodynamic  limit value of $\kappa$. In the case of NENP
$D\approx 0.18$ we find in the thermodynamic limit $\kappa_0\approx 2$.
The corresponding estimate for the in-plane anisotropy is $E\approx 0.012$.

We now take for granted the zero-field splittings and add a magnetic field
along the coordinate axis. We present the Lancz\H os results for the longest
chain we were able to deal with in figs.2-6. We find that the numerical points
are very well reproduced for all chain lengths by the following procedure:
we take the zero-field gaps as inputs in the perturbative formula (II-14)
and obtain the field behaviour. The corresponding curves are plotted
as solid lines in figs.2-6. One has of course to vary the gaps with
the chain length but the hyperbolic behaviour holds for {\it all} lengths.
The deviations between the exact results and the perturbative curves
are of the order of $10^{-2}$. In figs 2,3,4 the field lies respectively
along Z, X and Y. One of the gaps is barely affected and the other two
are split according to the simple formula (II-14). If ${\bf H}$ lies along
Z there is a crossing between the $\Delta_z$ and $\Delta_x$ modes (fig.2), if
${\bf H}$ lies along Y then there is a crossing between $\Delta_x$ and
$\Delta_y$ (fig.4).

It is interesting to note that, before the critical
field is reached, one observes that other states with higher spin begin
to arrive from higher energies. This is seen in all our figures:
the upper "triplet" mode always follow the perturbative trend but there
are new states that are lower in energy close to $H_c$ that do not belong to
the
Haldane triplet. These states will ultimately cross the ground state after
the transition point$^{19}$.

Let us consider the case of fig.2 where ${\bf H}$ lies along Z. Then
the operation $R^{\pi}_z$ is a symmetry operation of the Hamiltonian.
We can classify the triplet members according to their behaviour under
$R^{\pi}_z$:
$$
R^{\pi}_z |z\rangle = + |z \rangle ,
\eqno(III-1)
$$
while
$$
R^{\pi}_z |x\rangle = - |x \rangle \quad {\rm and}\quad
R^{\pi}_z |y\rangle = - |y \rangle .
\eqno(III-2)
$$
Thus no matrix element can avoid the crossing between these two members of the
triplet (x and z).
It is important to realize the following: when ${\bf H} =0$ and ${\cal H}_S =0$
we can demonstrate (III-1,2) for the degenerate triplet. By continuity
the behaviour under the preserved symmetry operations $R^{\pi}_{\alpha}$
will survive the addition of ${\bf H}$ and ${\cal H}_S$. This does not
rely upon perturbation theory but rather on the continuity of a discrete
quantum
number as a function of field and anisotropy. This reasoning holds also for
X and Y axis.

When the field no longer lies  along a coordinate axis, the crossings are
avoided.
A typical example is given in fig. 5. Here the field is very close to the Z
axis and thus there is an avoided crossing between the upper mode (z) and the
intermediate (x) mode.
This should be compared with fig. 2 (${\bf H} // Z$).
The solid lines in fig. 5 are obtained from the diagonalization
of perturbation (II-13). It is always very close to our Lancz\H os results.
We have checked that for various field arrangements the same property is true.
Another example is given in fig. 6 where the field lies close to the Y axis.

\bigskip
\noindent{\bf IV. CONCLUSION}
\medskip
We have studied the magnetic field behaviour of a realistic spin-1 chain
with the most general single-ion anisotropy that has proved adequate for NENP.
 The three gaps have been computed as
a function of the field by means of a Lancz\H os technique for chains of
lengths up to 16 spins. Our results are nicely reproduced by a simple
perturbation theory.
This perturbative approximation is
identical to the result of the fermionic effective theory$^{14}$ when
$K=\pi$. We have shown that one needs only to know
first-order perturbation theory to derive it in a satisfactory manner.
For arbitrary field orientation
and arbitrary chain length we have observed that the perturbative behaviour
holds. We thus infer that the
thermodynamic limit will be also described by the very same approximation.
Our {\it ab initio} results confirm some aspects of the effective theories
that have been applied to the spin-1 chain.

The field splitting of the Haldane gaps in figs. 2,3,4 is that found
in experiments on NENP$^{7-11}$. Both ESR and neutron scattering experiments
have observed the same splitting of the Haldane gap. Below the critical field
there is clearly a very good agreement between theory and experiments.
At large field the asymptotic behaviour of the wavefunctions leads to
a polarization of the modes which is similar to experimental data.

In the future it would be interesting to obtain the magnetization curves
in a realistic spin-1 following the lines of ref. 19. Another possibility
would be to investigate in detail the dynamical properties
under a field as has been done recently in the zero-field case$^{20,21}$.
\bigskip
\bigskip
\noindent
{\bf ACKNOWLEDGEMENTS}
\bigskip
\bigskip
We thank L. P. Regnault for informing us about experimental results on NENP.
Thanks are also due to J. Miller for reading our manuscript.
\vfill
\eject
\centerline{\bf REFERENCES}
\bigskip
\item{[1]}F. D. M. Haldane, Phys. Rev. Lett. {\bf 50}(1983) 1153; Phys. Lett.
A {\bf 93}(1983) 464.
\medskip
\item{[2]}For a recent review see: I. Affleck, J. Phys. Cond. Matter {\bf 1}
(1989) 3047.
\medskip
\item{[3]}W. J. L. Buyers, R. M. Morra, R. L. Armstrong, P. Gerlach and
K. Hirakawa, Phys. Rev. Lett. {\bf 56}, 371 (1986); R. M. Morra, W. J. L.
Buyers, R. L. Armstrong and K. Hirakawa, Phys. Rev. B{\bf 38}, 543 (1988).
\medskip
\item{[4]}A. Meyer, A. Gleizes, J.-J. Girerd, M. Verdaguer and O. Kahn,
Inorg. Chem. {\bf 21}, 1729 (1982).
\medskip
\item{[5]}J. P. Renard, M. Verdaguer, L. P. Regnault, W. A. C. Erkelens,
J. Rossat-Mignod and W. G. Stirling, Europhys. Lett. {\bf 3}, 945 (1987).
\medskip
\item{[6]}J. P. Renard, M. Verdaguer, L. P. Regnault, W. A. C. Erkelens,
J. Rossat-Mignod, J. Ribas, W. G. Stirling and C. Vettier, J. Appl. Phys.
{\bf 63}(8), 3538 (1988).
\medskip
\item{[7]}K. Katsumata, H. Hori, T. Takeuchi, M. Date, A. Yamagishi
and J. P. Renard, Phys. Rev. Lett. {\bf 63}, 86 (1989).
\medskip
\item{[8]}Y. Ajiro, T. Goto, H. Kikuchi, T. Sakakibara and T. Inami,
Phys. Rev. Lett. {\bf 63}, 1464 (1989).
\medskip
\item{[9]}L. P. Regnault, C. Vettier, J. Rossat-Mignod and J. P. Renard,
Physica B{\bf 180-181}, 188 (1992).
\medskip
\item{[10]}L. C. Brunel, T. M. Brill, I. Zalizniak, J. P. Boucher and J. P.
Renard, Phys. Rev. Lett. {\bf 69}, 1699 (1992).
\medskip
\item{[11]}W. Lu, J. Tuchendler, M. von Ortenberg and J. P. Renard,
Phys. Rev. Lett. {\bf 67}, 3716 (1991).
\medskip
\item{[12]}I. Affleck, Phys. Rev. B{\bf 41}, 6697 (1990); B{\bf 43}, 3215
(1991).
\medskip
\item{[13]}I. Affleck, Phys. Rev. B{\bf 46}, 9002 (1992).
\medskip
\item{[14]}A. M. Tsvelik, Phys. Rev. B{\bf 42}, 10499 (1990).
\medskip
\item{[15]}L. A. Takhtadjian, Phys. Lett. A{\bf 87}, 479 (1982);
H. M. Babudjian, Nucl. Phys. B{\bf 215}, 317 (1983).
\medskip
\item{[16]}O. Golinelli, Th. Jolic\oe ur and R. Lacaze, Phys. Rev. B{\bf 45},
9798 (1992).
\medskip
\item{[17]}O. Golinelli, Th. Jolic\oe ur and R. Lacaze, Phys. Rev. B{\bf 46},
10854 (1992).
\medskip
\item{[18]}L. P. Regnault, J. Rossat-Mignod and J. P. Renard, J. Mag. Mag.
Mater, {\bf 104}, 869 (1992).
\medskip
\item{[19]}T. Sakai and M. Takahashi, Phys. Rev. B{\bf 43}, 13383 (1991).
\medskip
\item{[20]}S. Ma, C. Broholm, D. H. Reich, B. J. Sternlieb and R. W. Erwin,
Phys. Rev. Lett. {\bf 69}, 3571 (1992).
\medskip
\item{[21]}O. Golinelli, Th. Jolic\oe ur and R. Lacaze, J. Phys. C.:
Condens. Matter {\bf 5}, 1399 (1993).
\medskip
\vfill
\eject
\centerline{\bf FIGURE CAPTIONS:}
\vskip 24pt
\bigskip
\bigskip
\noindent
{\bf Figure 1:}

\noindent
The coefficient $\kappa_0$ as a function of the anisotropy D up to D=0.5.
The raw data coming from the finite chain calculation are plotted as open
symbols: from bottom to top, N=4,6,8,10,12,14,16 and 18.
We have performed an extrapolation to the thermodynamic limit using
Shanks algorithm. The corresponding results are plotted as filled diamonds.
\bigskip
\bigskip
\noindent
{\bf Figure 2:}

\noindent
The three gaps in units of J as function of the magnetic field applied
along the Z axis. The points are results from Lancz\H os for a 16 spins
chain. The solid line is the result of perturbation theory eq.(II-14).
The dashed lines are the asymptotes of the hyperbola (II-14). Note the
crossing between the x and z modes. The two points on the right that deviate
seriously from the perturbative curves are in fact the energies of a
state that do not belong to the Haldane triplet and that has crossed
the upper members of the triplet. For the triplet state the deviations
always stay small.
\bigskip
\bigskip
\noindent
{\bf Figure 3:}

\noindent
The three gaps in units of J as function of the magnetic field applied
along the X axis. The symbols have the same meaning as in fig.2. There is
no crossing of levels before the critical field.
\bigskip
\bigskip
\noindent
{\bf Figure 4:}

\noindent
The three gaps in units of J as function of the magnetic field applied
along the Y axis. The symbols have the same meaning as in fig.2. Note the
crossing between the x and y modes.
\bigskip
\bigskip
\noindent
{\bf Figure 5:}

\noindent
The three gaps in units of J as function of the magnetic field applied
along a direction close to the Z axis,
defined by polar angles $\theta = 5$ and $\phi = 45$ (in degrees).
There is now an avoided crossing
between z and x modes since there is no longer a discrete symmetry to
allow a degeneracy. This is generic behaviour since the field is not
in a symmetry plane.
\bigskip
\bigskip
\noindent
{\bf Figure 6:}

\noindent
The three gaps in units of J as function of the magnetic field applied
along a direction close to the Y axis ($\theta = 85$ and $\phi = 80$).
The avoided crossing takes place
between x and y modes.
\bigskip
\bigskip
\medskip
\vfill
\eject
\nopagenumbers

\hfuzz=5pt
\baselineskip 12pt plus 2pt minus 2pt
\centerline{\bf MAGNETIC FIELD BEHAVIOUR OF A}
\centerline{\bf HALDANE GAP ANTIFERROMAGNET}
\vskip 24pt
\centerline{O. Golinelli,
Th. Jolic\oe ur,\footnote{*}{C.N.R.S. Research Fellow} and
R. Lacaze$^*$}
\vskip 12pt
\centerline{\it Service de Physique Th\'eorique\footnote{**}
{\rm Laboratoire de la Direction des Sciences de la Mati\`ere
 du Commissariat \`a l'Energie Atomique}}
\centerline{\it C.E.  Saclay}
\centerline{\it F-91191 Gif-sur-Yvette CEDEX, France}
\vskip 48pt
\vskip 24pt
\vskip 1.0in
\centerline{Submitted to: {\it J. Phys. C. Condensed Matter}}
\vskip 2.8in
\noindent April 1993

\noindent PACS No: 75.10J, 75.50E. \hfill SPhT/93-043
\vfill
\eject
\bye